# Comprehensive Numerical Hydrodynamic Analysis of Submarine in a Straight Course Simulation Using Wall-Resolved RANS Models

**N.Z. ABIDIN[a*], F. GRONDIN[a], P. MULLER[b], J.F. SIGRIST[c]**

a. Research Institute in Civil and Mechanical Engineering (GEM), Ecole Centrale de Nantes,
email : noh.bin-zainal-abidin@ec-nantes.fr, frederic.grondin@ec-nantes.fr
b. Sirehna, Naval Group (pol.muller@sirehna.com)
c. Naval expert, eye-pi (jfsigrist@wanadoo.fr)

## Résumé :

*Cette étude explore plusieurs facteurs clef affectant la précision de la prédiction par CFD de performances hydrodynamiques d'un sous-marin, et s'appuie sur des travaux antérieurs sur le maillage et l'analyse comparative des solveurs. Un modèle de sous-marin à échelle réduite est analysé numériquement à l'aide du modèle de turbulence RANS (Reynolds-Averaged Navier-Stokes) à un nombre de Reynold (Re) de $3,6 \times 10^6$ avec des maillages résolus aux parois ciblant un y+ <5. Sur la base de travaux antérieurs utilisant un maillage de $13 \times 10^6$ cellules généré avec SnappyHexMesh, une étude de convergence en maillage a été réalisée avec une résolution croissante de $15 \times 10^6$ cellules, $18 \times 10^6$ cellules, et $22 \times 10^6$ cellules. Cinq modèles de turbulence, à savoir k-ω SST, k-ω 2006, Lien Cubic, k-ε Launder-Sharma et Spalart-Allmaras, ont été évalués en termes de performance prédictive et d'efficacité de calcul. La représentativité de la couche limite a été examinée en comparant les profils de vitesse issus des maillages générés par l'outil de Cadence et par SnappyHexMesh à la loi théorique de la paroi. Dans la dernière partie, l'influence des appendices sur la résistance locale et globale a été analysée. Les résultats ont montré que le maillage le plus fin ($22 \times 10^6$ cellules) a permis d'obtenir une solution quasi indépendante du maillage avec une erreur de 1,16% et extrapolée à moins de 0,11%, ce qui est en accord avec les données expérimentales. Parmi les modèles de turbulence, le modèle k-ω SST a démontré la meilleure représentativité, avec une erreur sur les prévisions de résistance inférieure à 1,2 %. En outre, l'analyse de la loi de paroi a montré que le mailleur de Cadence permettait une résolution complète de la sous-couche visqueuse, tandis que la représentativité des calculs avec le maillage issu de SnappyHexMesh étaient limités par les dimensions des cellules normales aux parois. L'analyse des appendices a révélé que le massif seul contribuait jusqu'à 7,1 % de la traînée de pression et 7,7 % de la traînée visqueuse, tandis que l'appareil à gouverner en ajoutait respectivement 3,7 % et 5,6 %, ce qui implique un impact substantiel sur la résistance globale. Ces résultats contribuent au développement de méthodes CFD robustes et précises pour la résolution de l'hydrodynamique des sous-marins et offrent un cadre fiable pour l'évaluation future de la résistance dans le cadre de la conception d'applications maritimes complexes.*

## Abstract :

*This research explores several critical factors affecting CFD-based prediction accuracy of submarine hydrodynamics and builds upon previous work on preliminary mesh and solver benchmarking. A scaled*

*submarine model is analyzed numerically using the Reynolds-Averaged Navier-Stokes (RANS) turbulence model at a Reynold number (Re) of $3.6 \times 10^6$ with wall-resolved meshes targeting $y+ < 5$. Based on prior work using a $13 \times 10^6$ cells mesh generated with SnappyHexMesh, a mesh convergence study was performed with increasing resolution of fine ($15 \times 10^6$ cells), finer ($18 \times 10^6$ cells), and finest ($22 \times 10^6$ cells). Subsequently, five turbulence closure models which are k-ω SST, k-ω 2006, Lien Cubic, k-ε Launder-Sharma, and Spalart-Allmaras were assessed for predictive performance and computational efficiency. Boundary layer fidelity was examined by comparing velocity profiles extracted from Cadence and SnappyHexMesh meshes against the theoretical law of the wall. In the final phase, the influence of appendages on local and global resistance was analyzed. The results presented that the finest mesh ($22 \times 10^6$ cells) achieved a near mesh-independent solution with an error of 1.16% and extrapolated to less than 0.11% showing strong agreement with lab-scale data. Among the turbulence models, the k-ω SST model demonstrated the most reliable performance, with resistance predictions less than 1.2% of error. In addition, the law-of-the-wall analysis illustrated that the Cadence allowed full resolution of the viscous sublayer, while SnappyHexMesh performance was constrained by wall-normal cell sizing. Appendage analysis revealed that the sail alone contributed up to 7.1% of the pressure drag and 7.7% of viscous drag, while the rudder added 3.7% and 5.6% respectively, indicating a substantial impact on overall resistance. These findings contribute to developing robust and accurate CFD strategies for submarine hydrodynamics and offer a reliable framework for future resistance evaluation and design assessment in complex maritime applications.*

**Keywords: Submarine, CFD, Numerical Hydrodynamic, Straight Ahead, Wall Resolved RANS, Global resistance**

# 1    Introduction

Submarines have existed since before the Industrial Revolution and continue to evolve with technology [1]. Their changing shapes impact hydrodynamic performance, which is crucial for speed, power, and stealth [2]. Estimating propulsive power is essential in preliminary design stages. CFD enables accurate prediction and validation against experiments, supporting performance optimization. Turbulence and boundary layer modelling remain key challenges in submarine CFD, directly impacting drag and hydrodynamic performance. RANS models are commonly used for their balance of accuracy and efficiency [3], but reliable results depend on proper closure model selection and wall resolution [4,5]. Additionally, mesh quality plays a critical role in result accuracy [6]. This study was extended using the CFD model developed in Abidin et al. [7] by conducting a comprehensive numerical hydrodynamic analysis of a scaled submarine hull using RANS-based turbulence models with a wall-resolved mesh at Re of $3.6 \times 10^6$. The submarine model SSK class attack BB2 in model lab scale (1:35.1) utilized [8] and [9]. The MARIN provided the 3D CAD of a full-scale submarine, while Sirehna-Naval Group provided the lab-scale data based on the NATO AVT-301 collaboration project for validation purposes. This research used HPC resources of the Nautilus, GLiCID Computing Facility (Ligerien Group for Intensive Distributed Computing, https://doi.org/10.60487/glicid, Pays de la Loire, France) have been leveraged for simulations using high mesh resolution throughout this research.

# 2    Computational Setup

The finite volume method employed for discretizing the computational domain of the Navier-Stokes (NS) equation into small control volumes. All the solution fields will be stored in the centroid of the control volume. The segregated solver is utilized to solve the scalar matrix equations in an iterative sequence. This research utilises the RANS turbulence model of a wall-resolved mesh (y+ < 5). The numerical parameter will be configured in open-source CFD code, OpenFoam11 and post-processing in Paraview. The numerical simulation was conducted in a fully submerged submarine with isovolume and isothermal assumptions to simplify the condition. Reynolds decomposition splits the instantaneous variable $\phi(x,t)$ into mean $\bar{\phi}(x)$ and fluctuating $\phi'(x)$ components to form RANS equations. This introduces the Reynolds stress tensor, $\boldsymbol{R^\tau} = -\rho(\overline{\boldsymbol{u'u'}})$ from averaging the NS equation. $\boldsymbol{R^\tau}$ can be solved either by using direct computation corresponding to the six new equations with the symmetrical tensor or by referring to the Boussinesq hypothesis. The boundary and initial condition setup are as stated in [7]. The Semi-Implicit Method for Pressure-Linked Equations (SIMPLE) algorithm was utilised. To avoid the numerical diffusivity in the solution, the second-order accuracy is utilized for convection (bounded Gauss linear upwind) and diffusion (central differencing) terms in the NS equation. All schemes and solvers utilized are stored in the *fvScheme* and *fvSolution* code of OpenFOAM. Since this case is incompressible and has a steady state condition (RANS), the transient term is set to be a steady state. The simulation iteration is set to 1000 – 10,000 iterations to ensure convergence. The BB2 submarine had been scaled down to (1:35.1) to ensure the similarity towards the Lab-scale data provided by Sirehna, Naval Group. The BB2 submarine particulars are as in Table 1.

Table 1. BB2 Submarine particulars

| Description | Symbol / unit | Full (1:1) | Model (1:35.1) |
|---|---|---|---|
| Length overall | L (m) | 70.2 | 2 |
| Beam | B (m) | 9.6 | 0.2735 |
| Depth (to deck) | D (m) | 10.6 | 0.3020 |
| Depth (to top of sail) | Ds (m) | 16.2 | 0.4615 |

The BB2 submarine was rescaled down (1: 35.1) using CAD, Rhino and fixing the unattached surface to ensure the watertight condition. The 3D CAD converted to STL to suit the OpenFOAM meshing tool, SnappHexMesh and Cadence as illustrated in Fig.1. The free stream velocity, $U_r$=21 knots corresponding to the real scale applied, referring to Overpelt et al. [8] and Abidin et al. [7]. Thus, based on Froude similitude, ITTC [10], the $U_m = U_s/\sqrt{\lambda}$ had rescale down to $U_m = 1.8235 m/s$.

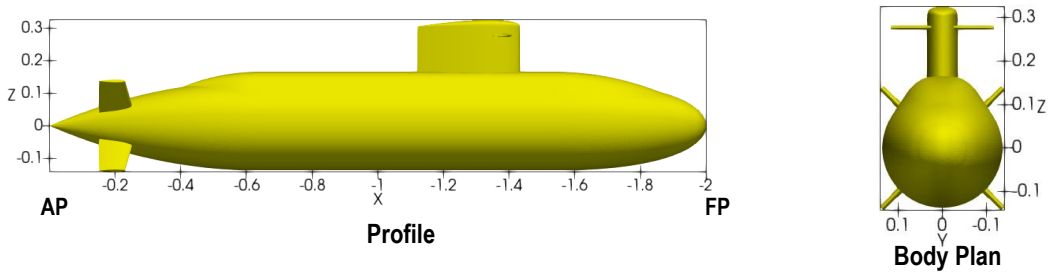

Fig.1: BB2 Submarine Geometry Preparation (Body and Profile View)

Initially, the RANS turbulence closure ($k - \omega SST$) employed for conducting mesh convergence studies following the procedures outlined by Celik et al. [11], three wall-resolved meshes were generated using Cadence, with a refinement ratio ($r$) of $\sqrt{2}$ at fine (15.7×10$^6$ cells), finer (18.4×10$^6$ cells) and finest (22.6×10$^6$ cells). Based on prior study, the optimum size of computational domain utilized was 1$L$ (front) x 1$L$ (lateral) x 3$L$ (wake). One refinement region is made near to wall with ratio of $B/L$=0.25 and $D/L$ =0.325 and the wake is 2.5L behind the submarine as shown in Fig.3. After optimizing the wall-resolved mesh and domain size, the study proceeded to evaluate various RANS turbulence closure models. The models are evaluated for their accuracy in predicting eddy viscosity and global hydrodynamic forces. Five turbulence models were evaluated: *k-ω SST* for accurate near-wall capture and adverse pressure gradient response [12]; *k-ω 2006* for improved transition modelling [13]; *Spalart-Allmaras* for efficient external flow prediction (Spalart et al., 1994); *k-ε Launder-Sharma* for robust boundary layer results with moderate cost (Launder et al., 1974); and *k-ε Lien Cubic*, a nonlinear model with the highest computational cost due to its treatment of nonlinear terms (Lien et al., 1996). Building on the optimized turbulence model, the study compares the law of the wall using the current optimized mesh against a prior snappyHexMesh case ($13\times10^6$ cells) [7] at streamwise locations $x/L$ =0.5, 0.4, and 0.35, as shown in Fig.3(a). Eventually the study extended to assess the impact of sail and rudder appendages on submarine resistance compared to the bare hull configuration [7] as shown in Fig.3(b).

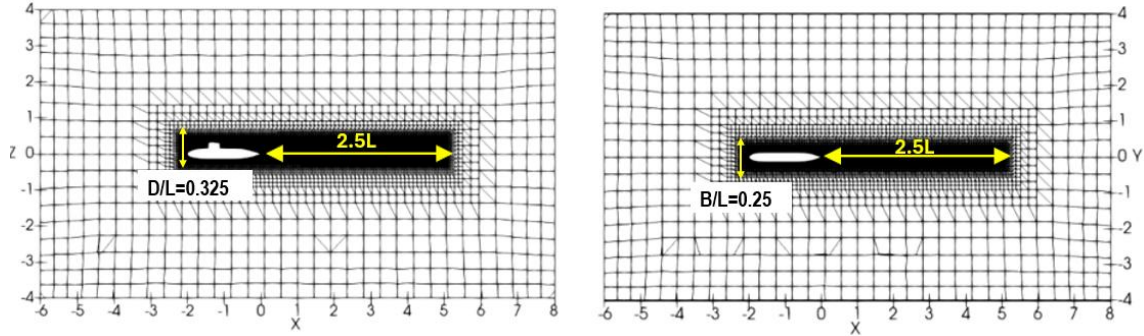

Fig.2: Refinement zone dimension (Profile and Top View). $22\times10^6$ cells (Cadence)

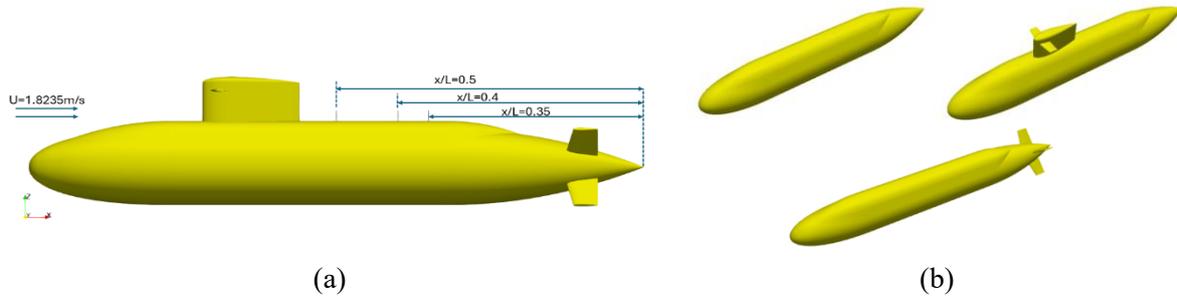

(a)                                                                  (b)

Fig.3: (a) Location of probes for measurement of Law of the Wall on hull deck, (b) hull configuration: bare hull, appendages (sail), appendages (rudder)

## 3.2   Results and Discussions

Fig.4 exhibits the distribution of $y^+< 5$, which ensures accurate resolution in the sub-viscous layer, indicates that a wall-resolved mesh was utilized. As in Table 2, the finest mesh ($22\times10^6$ cells) shows good convergence with a relative error, $e_{21}$ of 0.96% and $CGI_{21}$ of 1.42%, indicating low discretization uncertainty. A convergence ratio, $s$=1, confirms monotonic convergence [14,7]. Refinement ratios, $r_{21}$ and $r_{32}$ exceed 1.3, validating mesh reliability. The observed order of convergence $p = 1.76$ is close to ideal. Richardson extrapolation estimates global resistance as $F_0$=12.11N. $F_0$ can be converted to non-dimensional as $X_0 = F_0/(0.5\rho U^2 L^2)$ and compare to database (total force) of $X_{te}$=0.00182. As shown in Table 3, mesh error decreases with refinement, and Table 4 shows the extrapolated $F_0$ improves accuracy to 0.11%, compared to 3.11% in the prior study [7]. Fig.5(a) and (b) illustrate the comparison of five turbulence closure models towards $C_p = (p - p_\infty)/(0.5 \cdot \rho \cdot U^2)$ and $C_f = \tau/(0.5 \cdot \rho \cdot U^2)$ along the bottom of submarine hull. From x/L= 0 - 0.1 (rudder), high $C_p$ indicates stagnation and low-speed flow, with pressure drops due to separation and wake effects. All models show good agreement with prior studies [15,16]. Rapid pressure decline is observed between x/L= 0.1 - 0.3 due to flow acceleration over curved surfaces, increasing drag. From x/L= 0.3 - 0.9, the flow remains attached with uniform $C_p$. At the bow (x/L= 0.9 – 1.0), all models capture the pressure rise well; *k-ε Launder-Sharma* and *k-ε Lien Cubic* predict slightly higher peaks. For $C_f$, good agreement is seen across models. From x/L= 0 to 0.2, results align with prior studies. From x/L= 0.2 - 0.8, the flow is fully attached with consistent skin friction. At the bow, $C_f$ drops sharply near the stagnation point. *k-ε Launder-Sharma* slightly overpredicts wall shear, while the rest models provide better accuracy, especially in attached flow regions. All turbulence models achieved acceptable residuals: velocity below $10^{-5}$ and turbulence variables between $10^{-6}$ to $10^{-7}$. However, *Spalart-Allmaras* and *k-ε Launder-Sharma* showed slower or oscillatory convergence. *k-ω SST* and *k-ω 2006* were the most stable. First-order bounded upwind schemes on the divergence term were used for *Spalart-Allmaras* and *k-ε Launder-Sharma* to stabilize and improve convergence. Fig.6(a) shows that the 13×10⁶ cell mesh [7] at x/L = 0.5, 0.4, and 0.35 deviates from theoretical wall laws (Spalding, log-law) due to poor near-wall resolution (average $y^+$=7.2). In the sub-viscous layer ($y^+<5$), all profiles fail to follow $u^+= y^+$, and in the log-layer ($y^+ > 30$), underprediction is observed especially at x/L= 0.35 (purple) due to coarse or non-uniform mesh. In contrast, Fig.6(b) (15×10⁶ cell mesh from Cadence) shows excellent alignment with Spalding Law at all probes, confirming accurate near-wall resolution and stable boundary layer development. The improved first-layer thickness and prism layers enable better skin friction prediction, flow separation modelling, and turbulence reliability. Fig.7(a) compares local pressure ($F_p$) and viscous forces ($F_v$) between the prior 13×10⁶ cell mesh [7] and the refined 15×10⁶ cell wall-resolved (WR) mesh. The WR 15×10⁶ mesh accurately captures forces, especially near the nose, sail, and stern. At x/L = 0 – 0.2, both meshes show large negative $F_p$ due to adverse pressure gradients and possible flow separation, but WR 15×10⁶ mesh better resolves the pressure drop. $F_v$ is also higher in WR 15×10⁶, indicating improved wall shear prediction. Between x/L = 0.2 – 0.6, both forces stabilize; the 13×10⁶ mesh shows slight oscillations, while WR 15×10⁶ maintains higher $F_v$ due to finer near-wall resolution. At the sail region (x/L = 0.6), $F_p$ drops and $F_v$ peaks from sail interaction, with WR 15×10⁶ showing smoother, more realistic gradients. Near the bow (x/L = 0.8 – 1), $F_p$ rises sharply from stagnation, with 15×10⁶ capturing the buildup better, while $F_v$ briefly rises before dropping due to flow acceleration. While Fig.7(b) confirms a wall-resolved mesh generated with $y^+ < 5$ distribution along configurations. Fig.8 compares the local viscous force ($F_v$) along the hull for three configurations: bare hull (yellow), with rudder (red), and with sail (blue). Data from 50 longitudinal slices were integrated. At x/L = 0 – 0.2, the rudder case shows a sharp increase in $F_v$ due to high shear near the rudder root and leading edge. While bare and sail cases remain low. In the mid-body (x/L = 0.2 – 0.6), all cases show a gradual and stable rise in $F_v$ (5–6 N), indicating a fully developed boundary layer. At x/L = 0.6 – 0.7, the sail case peaks (~15 N) from intensified shear at the sail–hull junction. Near the bow (x/L = 0.7 – 1), all configurations show a decline due to reduced surface area, with a slight rise for the sail from wake interaction. Subsequently, Fig.9 shows the local pressure force ($F_p$) distribution. At x/L = 0 – 0.2, the rudder case peaks (~30 N) then drops sharply due to stagnation and adverse pressure. Bare and sail cases remain minimal. In the mid-body (x/L = 0.2 – 0.6), all show flat, low $F_p$ as drag is mostly viscous. At the sail region (x/L = 0.6 – 0.7), the sail case drops to ~–35 N due to suction from flow separation. Near the bow (x/L = 0.7 – 1), all cases rise sharply from stagnation pressure, with similar magnitudes, confirming independence from appendages. Table 6 summarizes the appendage contributions towards global resistance using the finest mesh (1.16% error). The bare hull accounts for 68.9% of viscous drag ($F_v$) and 6.93% of pressure drag ($F_p$), confirming skin friction as the dominant force. The sail adds 7.7% to $F_v$ and 7.08% to $F_p$ shown balanced impact on both components due to its bluff geometry and large wetted surface area. Meanwhile, the rudder adds 5.6% and 3.72%, respectively. Combined, the appendages contribute 13.3% to $F_v$ and 10.8% to $F_p$, significantly altering the drag profile and increasing the total resistance experienced by the submarine.

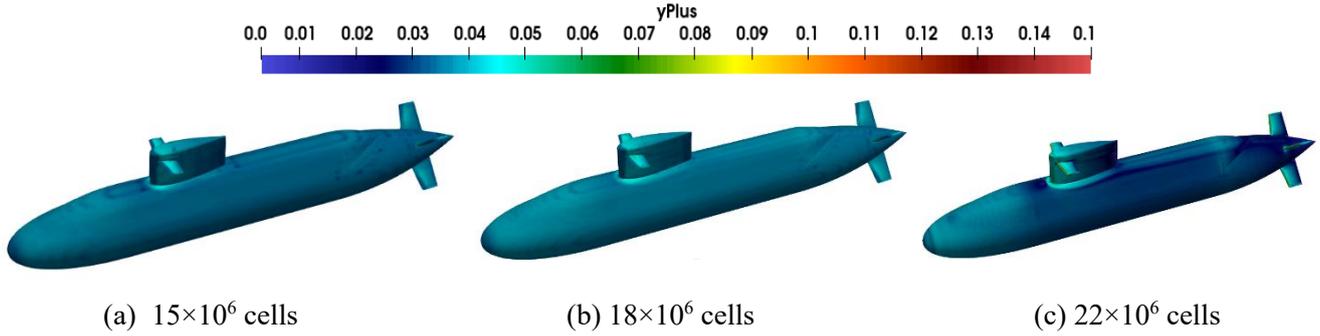

(a) 15×10⁶ cells     (b) 18×10⁶ cells     (c) 22×10⁶ cells

Fig.4: $y^+$ distribution along the submarine hull, Average $y^+=2$

Table 2. Model Outputs for Mesh Convergence Study

| $r_{21}(-)$ | $r_{32}(-)$ | $s\,(-)$ | $e_{21}(\%)$ | $e_{21}^{ext}(\%)$ | $p(-)$ | $CGI_{21}(\%)$ | $F_0$ (N) | $X_0$ |
|---|---|---|---|---|---|---|---|---|
| 1.415 | 1.413 | 1 | 0.96 | 1.15 | 1.76 | 1.42 | 12.11 | 0.001821 |

Table 3. Comparison of Global Resistance with Lab-scale data between HMR

| Mesh resolution | Base Size, $h$ (m) | No of Cells | $e_{exp}(\%)$ |
|---|---|---|---|
| Fine | 0.8 | 15,740,146 | 2.65 |
| Finer | 0.566 | 18,416,986 | 2.13 |
| Finest | 0.4 | 22,687,203 | 1.16 |

Table 4. Comparison of Extrapolated Global Resistance with Lab-scale data

| Author | Model | $F_0$ (N) | $e_{exp}(\%)$ |
|---|---|---|---|
| Abidin et al. [14] | RANS | 12.46 | 3.11 |
| Current (Finest Mesh) | RANS | 12.11 | 0.11 |

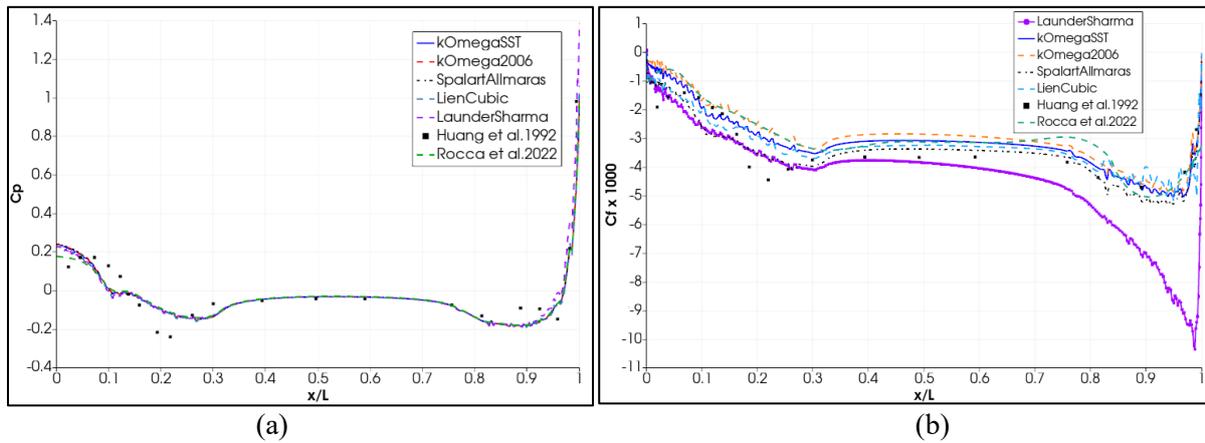

Fig.5: (a) Pressure Coefficient, $C_p$ (b) Skin Friction Coefficient, $C_f$ distribution over the bottom of the hull

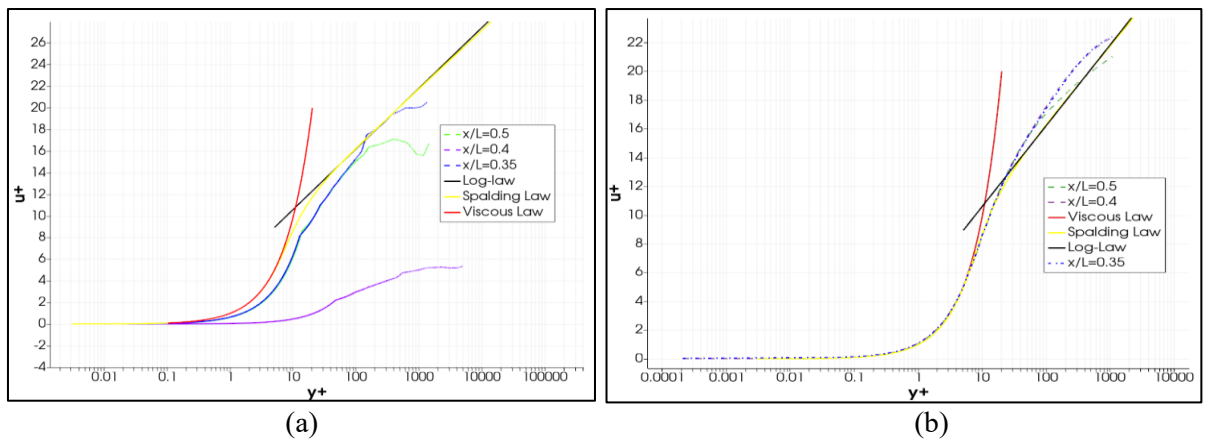

Fig.6: Law of the Wall (a) 13×10$^6$ cells (SnappyHexMesh) [7], (b) Wall-Resolved mesh, 15×10$^6$ cells (Fine-Table 3)

Table 5. Comparison of the Global Resistance of each turbulence closure model

| Turbulence Closure | $F_t$ (N) | $e_{exp}$(%) |
|---|---|---|
| *k-ω SST* | 12.24 | 1.2 |
| *k-ω 2006* | 11.55 | 4.5 |
| *Spalart-Allmaras* | 13.37 | 9.4 |
| *k-ε Lien Cubic* | 14.05 | 16.1 |
| *k-ε Launder-Sharma* | 15.03 | 24.2 |

Table 6. Percentage Contribution of Appendages towards Global Resistance

| Configurations | $F_p$ (%) | $F_v$ (%) |
|---|---|---|
| Appended (sail) | 7.08 | 7.7 |
| Appended (rudder) | 3.72 | 5.6 |
| Bare hull | 6.93 | 68.9 |
| **Total (%)** | **17.7** | **82.3** |

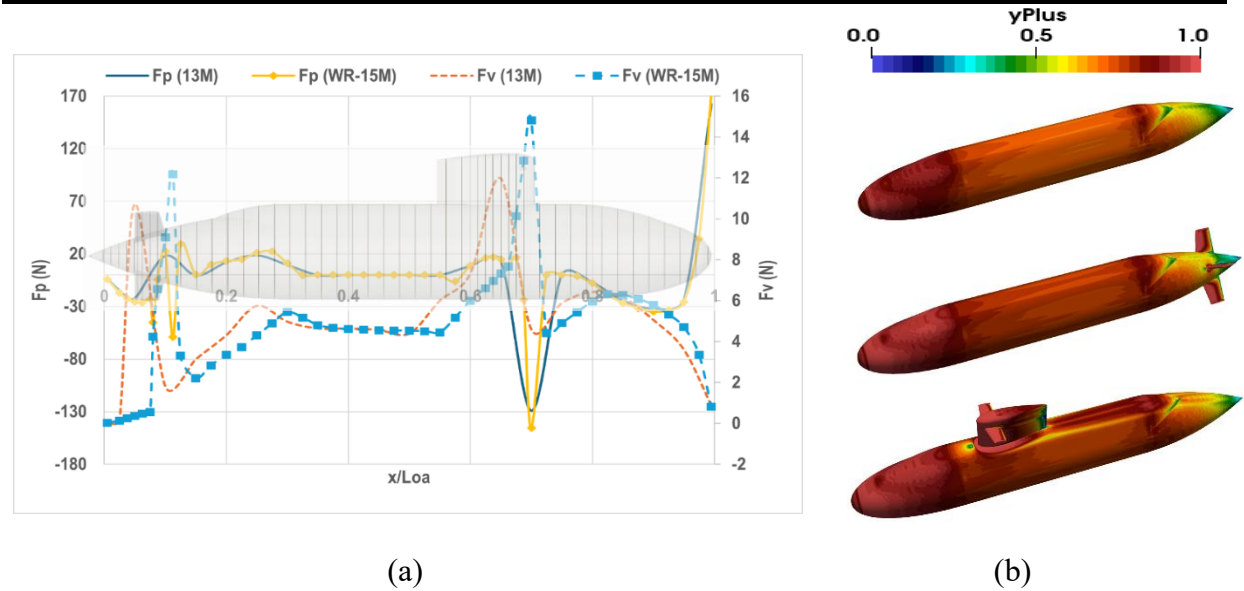

(a)                                                    (b)

Fig.7: (a) Comparison between Wall-resolved Mesh and Prior Mesh upon Local Forces, $F_v$ and $F_p$, (b) $y^+$ distribution along the submarine, bare hull-top, (b) appended (rudder)-mid, (c) appended (sail)-bottom

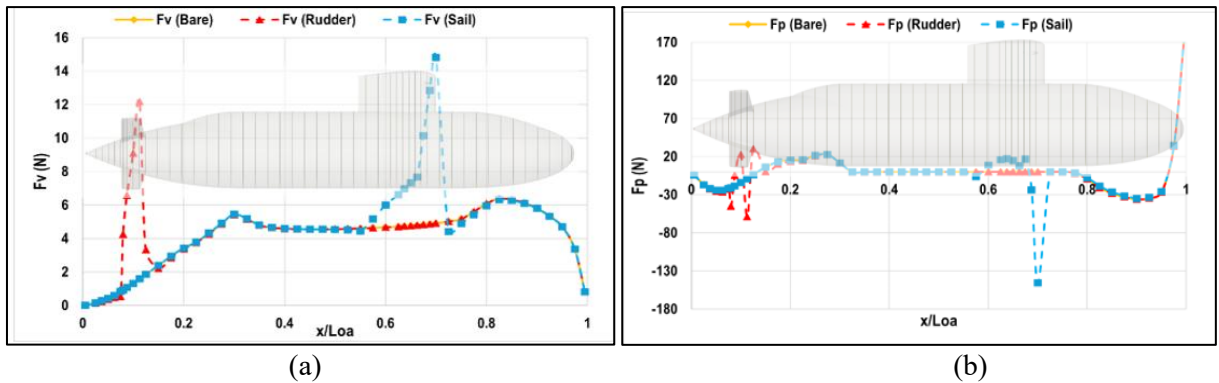

(a)                                                    (b)

Fig.8: (a) $F_v$ (b) $F_p$ distribution of bare hull, appended (rudder), appended (sail)

## 4.0   Conclusion

This study presents a comprehensive CFD-based investigation into the hydrodynamic performance of a submarine hull, corresponding to the mesh quality, turbulence closure modelling, and appendage contributions towards global resistance. The mesh convergence study across three wall-resolved grids confirmed that the finest mesh resolution improves solution accuracy with reduced discretization error to 1.16% compared to the previous, 3.11% [7] and GCI of 1.42% demonstrates an acceptable value, particularly for pressure and viscous force predictions. The optimized scale domain can be achieved with $β < 1\%$. A comparative evaluation of five RANS turbulence closure models revealed that the *k-ω SST* model provided the best balance between predictive accuracy and numerical stability *k-ω SST* with an error of 1.2%. Nevertheless, the models such as *k-ε Launder-Sharma* and *k-ε Lien Cubic* exhibited higher sensitivity and overpredicted form drag, which is incompatible with this case. Analysis of boundary layer behaviour using the law of the wall showed that the Cadence-generated mesh successfully resolved perfectly in the sub-viscous layer ($y^+ < 5$) and log-layer ($y^+ > 30$) whereas the SnappyHexMesh struggled to capture near-wall velocity gradients accurately. Nevertheless, the global error obtained by SnappyHexMesh is still acceptable, less than 5%. Eventually, appendage contribution analysis demonstrated that both the sail and rudder considerably modified local flow characteristics and contributed to the overall resistance, with the sail contributing a total drag of 14.78%. In comparison, the rudder contributed 9.32% towards global resistance. These findings emphasize the importance of high-fidelity meshing, domain sizing, turbulence model selection, and detailed appendage modelling in submarine resistance prediction to enhance the robust and accurate CFD approaches beneficial for preliminary and detailed industrial analysis in complex maritime applications. Future work should consider transitioning to hybrid RANS-LES models to capture unsteady wake dynamics, especially in the aft region where turbulence plays a dominant role towards overall hydrodynamic performance.

# Références